\documentclass[conference]{IEEEtran}
\IEEEoverridecommandlockouts

\usepackage{cite}
\usepackage{amsmath,amssymb,amsfonts}
\usepackage{mathtools}
\usepackage{bm}
\usepackage{hyperref}
\usepackage{algorithm}
\usepackage{algpseudocode}
\usepackage{graphicx}
\usepackage{textcomp}
\usepackage{xcolor}
\usepackage{braket}
\usepackage{qcircuit}
\usepackage{siunitx}   
\usepackage{booktabs}  
\usepackage{stfloats} 
\usepackage{lmodern}     
\usepackage[T1]{fontenc}  

\usepackage{tikz}
\usetikzlibrary{arrows.meta,positioning,fit,shapes.misc}

\usepackage{hyperref}
\hypersetup{
    colorlinks=true,
    linkcolor=blue,
    urlcolor=cyan,
}

\DeclarePairedDelimiter{\norm}{\lVert}{\rVert}
\DeclarePairedDelimiter{\inner}{\langle}{\rangle}

\usepackage{atbegshi}
\usepackage{eso-pic}
\AtBeginShipout{%
  \ifnum\value{page}=4
\global\setbox\AtBeginShipoutBox=\vbox{\vspace{2pt}\unvbox\AtBeginShipoutBox}%
  \fi
}
\usetikzlibrary{positioning, calc, shapes.geometric, arrows.meta}

\def\BibTeX{{\rm B\kern-.05em{\sc i\kern-.025em b}\kern-.08em
    T\kern-.1667em\lower.7ex\hbox{E}\kern-.125emX}}

\begin{document}

\title{Quantum Reinforcement Learning-Guided \\Diffusion Model for Image Synthesis via Hybrid Quantum-Classical Generative Model Architectures

}

\author{
\IEEEauthorblockN{Chi-Sheng Chen}
\IEEEauthorblockA{\textit{Independent Researcher} \\
Cambridge, USA \\
m50816m50816@gmail.com}
\and
\IEEEauthorblockN{En-Jui Kuo}
\IEEEauthorblockA{\textit{Department of Electrophysics} \\
 National Yang Ming Chiao Tung University\\
 Hsinchu, Taiwan \\
kuoenjui@nycu.edu.tw}
}

\maketitle

\begin{abstract}
Diffusion models typically employ static or heuristic classifier-free guidance (CFG) schedules, which often fail to adapt across timesteps and noise conditions. In this work, we introduce a quantum reinforcement learning (QRL) controller that dynamically adjusts CFG at each denoising step. The controller adopts a hybrid quantum--classical actor--critic architecture: a shallow variational quantum circuit (VQC) with ring entanglement generates policy features, which are mapped by a compact multilayer perceptron (MLP) into Gaussian actions over $\Delta$CFG, while a classical critic estimates value functions. The policy is optimized using Proximal Policy Optimization (PPO) with Generalized Advantage Estimation (GAE), guided by a reward that balances classification confidence, perceptual improvement, and action regularization. Experiments on CIFAR-10 demonstrate that our QRL policy improves perceptual quality (LPIPS, PSNR, SSIM) while reducing parameter count compared to classical RL actors and fixed schedules. Ablation studies on qubit number and circuit depth reveal trade-offs between accuracy and efficiency, and extended evaluations confirm robust generation under long diffusion schedules.
\end{abstract}


\section{Introduction}
Diffusion models have become a leading paradigm for image synthesis, inpainting, and editing, reaching state-of-the-art quality through iterative denoising from a noisy latent $x_t$ toward a clean sample $x_0$ using a learned time-dependent score function or denoiser \cite{ho2020ddpm,song2020score}. Sampling can be accelerated via non-Markovian solvers such as DDIM, which preserve training objectives while reducing the number of function evaluations \cite{song2020ddim}. Beyond architectural and solver design, \emph{guidance} strongly impacts semantic faithfulness and diversity in conditional generation. Classifier guidance injects gradients from an external classifier into the reverse process \cite{dhariwal2021diffusion}, whereas Classifier-Free Guidance (CFG) combines conditional and unconditional predictions within a single model to provide a tunable guidance scale $w$ \cite{ho2022classifierfree}. In its common parameterization,
\begin{equation}
\hat{\epsilon}_{\theta}(x_t,c) \;=\; \epsilon_{\theta}(x_t,\varnothing) \;+\; w \bigl(\epsilon_{\theta}(x_t,c)-\epsilon_{\theta}(x_t,\varnothing)\bigr),
\label{eq:cfg}
\end{equation}
where $\epsilon_{\theta}$ is the noise predictor and $c$ denotes the conditioning signal.

Despite its central role, CFG is typically used with \emph{static or hand-crafted schedules} (e.g., constant $w$, linear or cosine annealing). Such heuristics struggle to adapt to the evolving generative state across timesteps, noise levels, and prompts. Empirically, the optimal $w_t$ depends on the instantaneous signal-to-noise ratio, model calibration, and intermediate sample quality. Recent work has begun to re-examine guidance design and training-free calibrations \cite{ma2024elucidating,sadat2024rethinking}, yet the prevailing practice remains schedule-based rather than state-aware.

\paragraph{Our perspective.}
We cast guidance scheduling as a \emph{sequential decision-making} problem and learn a lightweight controller with reinforcement learning (RL). Concretely, we treat the diffusion trajectory as an environment and adjust guidance \emph{incrementally} at each step via an action $a_t=\Delta\!\text{CFG}_t$ derived from compact state features (timestep, denoising residual statistics, and proxy-classifier feedback). The controller is trained to maximize a return that trades off perceptual quality and stability:
\begin{equation}
\begin{split}
J(\pi_\theta) \;=\; \mathbb{E}_{\pi_\theta}\!\left[\sum_{t=1}^{T} \gamma^{t-1} \, r_t \right], \quad \\
r_t \;=\; \alpha \cdot \text{Conf}_t \;+\; \beta \cdot \Delta \text{Qual}_t \;-\; \lambda \|a_t\|_2^2,
\label{eq:rlobjective}
\end{split}
\end{equation}
where $\text{Conf}_t$ is proxy classification confidence, $\Delta \text{Qual}_t$ measures stepwise improvement (e.g., LPIPS/SSIM proxies over denoised estimates), and the last term regularizes action magnitude. We optimize with Proximal Policy Optimization (PPO) and Generalized Advantage Estimation (GAE) for stable, on-policy learning \cite{schulman2017ppo,schulman2015gae}.

\paragraph{Hybrid quantum–classical control.}
To achieve parameter efficiency without sacrificing expressivity, we adopt a hybrid quantum–classical actor–critic. The actor uses a shallow variational quantum circuit (VQC) with ring entanglement to transform classical state features into quantum features; expectation measurements then feed a compact MLP head that outputs a Gaussian policy over $a_t=\Delta\!\text{CFG}_t$. The critic remains classical to stabilize value learning. Prior studies suggest VQCs can offer expressive function families with fewer trainable parameters and, in some settings, theoretical separations for policy classes \cite{chen2020vqc_drl,jerbi2021parampolicies,skolik2022quantumagents}. Closely related, deep RL has also been used to design quantum circuits (quantum architecture search), highlighting the synergy between RL and quantum representations \cite{kuo2021qas}. Complementary evidence from quantum-enhanced generative modeling in vision and language supports the potential of quantum features under tight capacity or low-data regimes \cite{chen2025qgm_image,chen2025qnlg}, while applied QRL case studies underscore the importance of reward design when aligning proxy signals with downstream objectives \cite{chen2025qrl_sector}.

\paragraph{Contributions.}
(i) We formulate diffusion guidance scheduling as RL and introduce a \emph{state-aware} controller that outputs per-step $\Delta\!\text{CFG}$; (ii) we propose a \emph{hybrid VQC–MLP actor} with a classical critic trained by PPO/GAE \cite{schulman2017ppo,schulman2015gae}; (iii) we design a compact state–action–reward interface that plugs into standard samplers (e.g., DDIM \cite{song2020ddim}); and (iv) we provide empirical evidence on CIFAR‑10 that dynamic control improves robustness and parameter efficiency relative to static schedules and classical capacity-matched actors. Our results complement contemporaneous guidance studies \cite{ma2024elucidating,sadat2024rethinking} by showing that \emph{learned} stepwise control can further exploit the evolving denoising dynamics.

\begin{table*}[!t]
\centering
\caption{Quality and efficiency on CIFAR-10 (class-conditional). Fill \textbf{TBD} with actual numbers.}
\label{tab:main}
\sisetup{detect-all}
\begin{tabular}{lcccccc}
\toprule
Model & PSNR$\uparrow$ & SSIM$\uparrow$ & LPIPS$\downarrow$ & Params(K)$\downarrow$  \\
\midrule
Classical RL (MLP) & 9.6454 ± 1.6090 & 0.0438 ± 0.0657 & 0.2221 ± 0.0722 & $\approx$9.3 \\
\textbf{QRL (VQC+MLP)} & \textbf{9.9456 ± 1.7202} & \textbf{0.0438 ± 0.0616} & \textbf{0.2142 ± 0.0650} & \textbf{$\approx$2.5}  \\
\bottomrule
\end{tabular}
\end{table*}

\section{Related Work}
\label{sec:related}

\textbf{Diffusion guidance and sampling control.}
Diffusion models have become a leading paradigm for high‑quality image synthesis. Classifier guidance and classifier‑free guidance (CFG) are now standard tools to trade off semantic fidelity against diversity in conditional generation. Classifier guidance uses gradients from an external classifier to bias the reverse process, while CFG combines conditional and unconditional scores to provide a tunable guidance scale $w$.\;DDIM further accelerates sampling by replacing the reverse Markov chain with a non‑Markovian, training‑equivalent process.\;However, most works apply \emph{fixed} or hand‑crafted guidance schedules (constant, linear, cosine), which are not adaptive to the evolving generative state, noise level, or prompt/label.\;We view guidance scheduling as a \emph{sequential decision} problem that benefits from step‑wise control rather than static heuristics~\cite{dhariwal2021diffusion,ho2022classifierfree,song2020ddim}.

\textbf{Reinforcement learning (RL) for generative control.}
Casting generation dynamics as an interactive environment allows a controller to optimize task‑aligned rewards while reacting to intermediate states. In practice, policy gradient methods such as Proximal Policy Optimization (PPO) and variance‑reduced estimators such as Generalized Advantage Estimation (GAE) provide stable, sample‑efficient updates for on‑policy training, which makes them attractive for lightweight, per‑step controllers that must operate inside a diffusion sampler~\cite{schulman2017ppo,schulman2015gae}. Yet RL specifically for \emph{guidance scheduling} in diffusion models remains underexplored; our formulation offers an \emph{plug‑in} controller that learns $\Delta\!\mathrm{CFG}$ from compact state summaries.

\textbf{Quantum RL and parameterized quantum policies.}
Quantum machine learning has proposed hybrid quantum‑classical policies based on variational quantum circuits (VQCs) for RL~\cite{chen2023qdrl}. Early demonstrations showed that VQCs can approximate value functions and policies with fewer trainable parameters, sometimes conferring representational benefits under tight capacity constraints. Subsequent work established parameterized quantum policies with theoretical separations on carefully constructed tasks and analyzed design choices (encodings, observables) that affect trainability; VQC‑based deep Q‑learning has also been validated on Gym benchmarks. Extensions to continuous action spaces introduce quantum DDPG variants that avoid coarse discretization. Separately, RL has proven effective for \emph{quantum architecture search}, automatically composing gate sequences. Our hybrid controller follows this line by using a shallow VQC to generate policy features and a small MLP head for Gaussian actions over $\Delta\!\mathrm{CFG}$, while a classical critic stabilizes learning~\cite{chen2020vqc_drl,jerbi2021parampolicies,skolik2022quantumagents,wu2025qrlcontinuous,kuo2021qas}.

\textbf{QML for generative modeling.}
Recent QML studies have begun to evaluate quantum‑enhanced generators in vision and language.\;For images, hybrid quantum–classical diffusion or bottlenecked generators report potential benefits in low‑data regimes (e.g., MNIST/MedMNIST) under capacity‑matched settings; for language, multi‑model frameworks compare quantum‑enhanced attention/RNN‑style variants to Transformers across metrics such as perplexity, BLEU, diversity, and repetition, finding niche advantages (e.g., diversity control) but also gaps in domain robustness.\;These results collectively motivate exploring \emph{parameter‑efficient} quantum features as policy components when learning compact, step‑wise controllers inside diffusion samplers~\cite{chen2025qgm_image,chen2025qnlg}.

Prior work shows (i) guidance is crucial but typically scheduled statically; (ii) RL provides a principled pathway to \emph{adaptive} control; and (iii) VQCs can supply expressive, low‑parameter features for policies.\;We integrate these strands by training a hybrid QRL controller that adjusts CFG \emph{per denoising step} to improve perceptual quality and efficiency.

\begin{figure}
    \centering
    \includegraphics[width=1\linewidth]{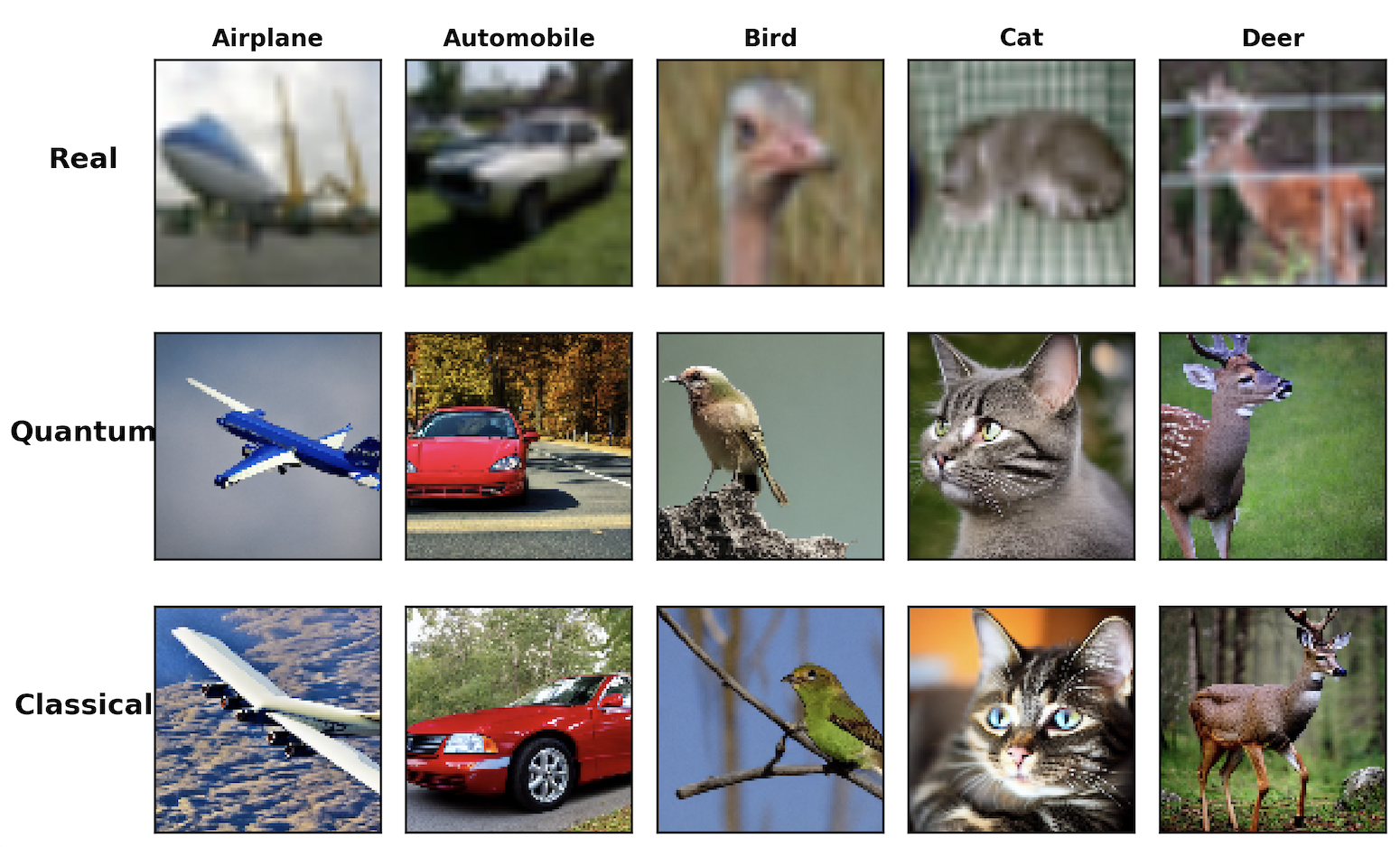}
    \caption{The comparison of generated images.}
    \label{fig:img_gen}
\end{figure}

\section{Methodology}

\subsection{Overview}
We control a pretrained diffusion sampler by adjusting CFG at each timestep. The policy $\pi_\theta(a_t\mid s_t)$ outputs $a_t\in[-2,2]$ and we set $\text{CFG}_t=\mathrm{clip}(\text{CFG}_0+a_t,\,1,\,12)$. Standard CFG combines conditional and unconditional noise predictions as
\begin{equation}
\hat\epsilon\,=\,\epsilon_{\text{uncond}}+g\,\big(\epsilon_{\text{cond}}-\epsilon_{\text{uncond}}\big),\quad g\triangleq \text{CFG}_t.
\end{equation}

\subsection{Dataflow Description}
The generation process starts from a natural language prompt (e.g., \emph{``a photo of a cat''}), which is encoded into a sequence embedding of dimension $[77,768]$ by the CLIP text encoder. Alongside this embedding, the environment state---comprising latent statistics, diffusion timestep, and semantic confidence---is represented as a compact vector of size $[6]$. This state is passed into the quantum actor, where a variational quantum circuit, followed by classical post-processing, produces an adaptive scalar value corresponding to the Classifier-Free Guidance (CFG) weight. The CFG weight modulates the conditioning strength of the text embedding before being injected into the denoising network. 

The latent variable is updated iteratively through denoising, and this process is repeated for $N$ timesteps until a clean latent code is obtained. The final latent representation is then decoded by a Variational Autoencoder (VAE) into an RGB image of resolution $[3,64,64]$.

To evaluate the generated sample, a frozen classifier computes semantic confidence, which is combined with auxiliary quality signals to produce a scalar reward. Finally, Proximal Policy Optimization (PPO) leverages this reward to update both the policy (quantum actor) and the value function, thereby closing the reinforcement learning loop.

\subsection{State, Action, Reward}
The state $s_t\in\mathbb{R}^6$ collects online features: normalized timestep $t/T$, latent norm $\norm{z_t}_2$, noise norm $\norm{\epsilon_t}_2$, dot product $\inner{z_t}{\epsilon_t}$, previous action $a_{t-1}$, and a proxy confidence $p_{\text{proxy}}(y\mid x_t)$. The action is $a_t=\Delta\!\text{CFG}\in[-2,2]$ with clipping described above. The reward is
\begin{align}
R_t &= \alpha\,R_{\text{cls}}(x_t) + \beta\,R_{\text{step}}(x_t, x_{t-1}) - \lambda_{\text{act}}\,\lVert a_t\rVert^2 - \lambda_{\text{tv}}\,\mathrm{TV}(x_t),\label{eq:reward}
\end{align}
with defaults $\alpha=1.0$, $\beta=0.2$, $\lambda_{\text{act}}=5\times10^{-3}$, $\lambda_{\text{tv}}=0$.

\subsection{Hybrid Quantum--Classical Actor}
We use angle encoding on each qubit: for qubit $i$, apply $\mathrm{RY}(\theta_i)\,\mathrm{RZ}(\phi_i)$. The VQC has $n_q{=}4$ qubits and depth $L{=}2$. Each layer composes Rotations, ring CNOT entanglement, and a final $\mathrm{RX}$ per qubit. Pauli-Z expectations on all qubits yield $q\in\mathbb{R}^{n_q}$, which a small MLP maps to $(\mu, \log\sigma)$. Actions are sampled via reparameterization $a_t=\mu+\exp(\log\sigma)\,\epsilon$ and clamped to $[-2,2]$.

\subsection{Critic and PPO}
The critic $V_\psi(s)$ is an MLP. We use GAE for advantages and PPO with clip $\epsilon=0.1$ and entropy/value losses (weights $c_H{=}0.01$, $c_v{=}0.5$). LRs: actor $10^{-4}$, critic $10^{-3}$; rollouts: $N{=}8$ envs, horizon $T{=}512$, epochs $K{=}4$, minibatch $B{=}8$.

\begin{algorithm}[t]
\caption{QRL--PPO for Dynamic CFG Control}
\label{alg:qrlppo}
\begin{algorithmic}[1]
\Require pretrained diffusion sampler components; base guidance $\text{CFG}_0$; hyperparams $(\gamma,\lambda,\epsilon,\alpha,\beta,\lambda_{\text{act}},\lambda_{\text{tv}})$
\State Initialize policy parameters $\theta$ (VQC+MLP), critic $\psi$, replay buffers
\For{iteration $=1$ to $I$}
\For{environment $=1$ to $N$} \Comment{parallel rollouts}
\State Reset env, sample prompt/text embedding, init latent $z_1$, set $a_0\gets0$
\For{$t=1$ to $T$}
\State Build state $s_t=[t/T,\norm{z_t},\norm{\epsilon_t},\inner{z_t}{\epsilon_t},a_{t-1},p_{\text{proxy}}]$
\State $(\mu,\log\sigma)\gets \pi_\theta(s_t)$; sample $a_t\sim\mathcal{N}(\mu,\sigma^2)$; $a_t\gets\mathrm{clip}(a_t,-2,2)$
\State $g_t\gets \mathrm{clip}(\text{CFG}_0 + a_t,\,1,\,12)$
\State $\hat\epsilon\gets \epsilon_{\text{uncond}} + g_t(\epsilon_{\text{cond}}-\epsilon_{\text{uncond}})$
\State $z_{t+1}\gets \textsc{SchedulerStep}(z_t,\hat\epsilon, t)$; decode image $x_{t+1}$ (for reward only)
\State $r_t\gets$ Eq.~\eqref{eq:reward}; store $(s_t,a_t,r_t,\log\pi_\theta(a_t\mid s_t), V_\psi(s_t))$
\EndFor
\EndFor
\State Compute advantages $\hat A_t$ via GAE$(\gamma,\lambda)$ and returns $\hat R_t$
\State Update $\theta$ by maximizing PPO objective with clip $\epsilon$ and entropy bonus $c_H$
\State Update $\psi$ by minimizing MSE to $\hat R_t$ with weight $c_v$
\EndFor
\end{algorithmic}
\end{algorithm}

\begin{algorithm}[t]
\caption{Inference with Trained QRL Controller}
\label{alg:inference}
\begin{algorithmic}[1]
\Require trained policy $\theta$, base guidance $\text{CFG}_0$, prompt embedding, steps $T$
\State Initialize latent $z_1\sim\mathcal{N}(0,I)$; set $a_0\gets 0$
\For{$t=1$ to $T$}
\State Build state $s_t$ from current sampler statistics and $a_{t-1}$
\State $(\mu,\log\sigma)\gets \pi_\theta(s_t)$; set $a_t\gets\mathrm{clip}(\mu, -2, 2)$ \Comment{deterministic mean action}
\State $g_t\gets \mathrm{clip}(\text{CFG}_0 + a_t,\,1,\,12)$
\State $\hat\epsilon\gets \epsilon_{\text{uncond}} + g_t(\epsilon_{\text{cond}}-\epsilon_{\text{uncond}})$; $z_{t+1}\gets \textsc{SchedulerStep}(z_t,\hat\epsilon, t)$
\EndFor
\State \Return $\textsc{Decode}(z_{T+1})$
\end{algorithmic}
\end{algorithm}

\section{Experiments}

\subsection{Experimental Setup}

\textbf{Dataset.} 
We employ the CIFAR-10 dataset, which consists of 60,000 natural images of size $32 \times 32 \times 3$ drawn from $10$ object classes. The dataset is split into $50{,}000$ training samples and $10{,}000$ test samples, denoted as 
\begin{equation}
\mathcal{D}_{\text{train}} = \{(x_i, y_i)\}_{i=1}^{50{,}000}, \quad 
\mathcal{D}_{\text{test}} = \{(x_j, y_j)\}_{j=1}^{10{,}000},
\end{equation}
where $x \in \mathbb{R}^{32 \times 32 \times 3}$ is an image and $y \in \{1,\dots,10\}$ is the class label. Images are normalized to the range $[-1,1]$ before being fed into the generative model.

\textbf{Sampler.} 
We adopt the Denoising Diffusion Implicit Model (DDIM), stable-diffusion-v1.5, with $50$ timesteps for efficient sampling. Given the forward noising process
\begin{equation}
q(x_t \mid x_{t-1}) = \mathcal{N}\!\left(\sqrt{\alpha_t} x_{t-1}, (1-\alpha_t) I \right),
\end{equation}
the DDIM sampler deterministically reconstructs $x_0$ via
\begin{equation}
x_{t-1} = \sqrt{\alpha_{t-1}} \hat{x}_0 + \sqrt{1-\alpha_{t-1}} \,\epsilon_\theta(x_t, t),
\end{equation}
where $\epsilon_\theta$ is the trained denoising network and $\hat{x}_0$ is the predicted clean image.

\textbf{Proxy Classifier.} 
A frozen ResNet-18, pretrained on CIFAR-10, is used as a proxy classifier to evaluate semantic fidelity. Formally, we use the classifier
\begin{equation}
f_\phi: \mathbb{R}^{32 \times 32 \times 3} \to \Delta^{9},
\end{equation}
where $\Delta^{9}$ is the $10$-class probability simplex, to provide an external semantic validation signal.

\textbf{Baselines.} 

We compare against: (i) a \emph{classical MLP actor} matched in parameter count to the quantum-classical actor for fair capacity comparison.

\textbf{Metrics.} 
We employ multiple quantitative metrics:
\begin{itemize}
    
    \item \textbf{Learned Perceptual Image Patch Similarity (LPIPS$\downarrow$)}: 
    \begin{equation}
    \text{LPIPS}(x, \hat{x}) = \sum_l \frac{1}{H_l W_l} \sum_{h,w} \| \hat{y}_{hw}^l - y_{hw}^l \|_2^2,
    \end{equation}
    where $y^l = \phi^l(x)$ are layer-$l$ feature activations.
    
    \item \textbf{Peak Signal-to-Noise Ratio (PSNR$\uparrow$)}:
    \begin{equation}
    \text{PSNR}(x,\hat{x}) = 10 \log_{10} \frac{L^2}{\text{MSE}(x,\hat{x})}, \quad L=255,
    \end{equation}
    where $\text{MSE}$ denotes mean squared error.
    
    \item \textbf{Structural Similarity Index (SSIM$\uparrow$)}:
    \begin{equation}
    \text{SSIM}(x, \hat{x}) = \frac{(2\mu_x \mu_{\hat{x}} + C_1)(2\sigma_{x\hat{x}}+C_2)}{(\mu_x^2 + \mu_{\hat{x}}^2 + C_1)(\sigma_x^2 + \sigma_{\hat{x}}^2 + C_2)},
    \end{equation}
    with $\mu, \sigma$ denoting mean and variance, and $\sigma_{x\hat{x}}$ cross-covariance.
\end{itemize}


\section{Results and Conclusion}
\label{sec:results}

Table~\ref{tab:main} summarizes the quantitative evaluation on CIFAR-10 in the class-conditional setting. We report four widely used metrics: peak signal-to-noise ratio (PSNR), structural similarity index (SSIM), learned perceptual image patch similarity (LPIPS), and model parameter count. 

Overall, the proposed quantum reinforcement learning (QRL) actor achieves superior performance compared to the classical MLP baseline. Specifically, QRL obtains a higher PSNR ($9.95$ vs. $9.65$) and lower LPIPS ($0.214$ vs. $0.222$), indicating both better reconstruction fidelity and improved perceptual quality. The SSIM score is also marginally better, suggesting more stable structural preservation across generated samples. Notably, these improvements are obtained with fewer parameters ($\approx 2.5$K vs. $\approx 9.3$K), highlighting the efficiency of hybrid quantum-classical representations.

Complementary to the quantitative metrics, Fig.~\ref{fig:img_gen} provides a visual comparison of generated images.  This figure provides a comprehensive visual comparison of image generation quality across three categories: real images (top row), quantum-generated images (middle row), and classical-generated images (bottom row), spanning ten object categories from CIFAR-10 (airplane, automobile, bird, cat, deer, dog, frog, horse, ship, and truck), each displayed at 512×512 resolution (The original CIFAR-10 images are 32×32, we upsampled in this figure). The quantum-guided model produces sharper textures and more semantically faithful objects compared to the classical baseline. These qualitative observations align with the numerical trends reported in Table~\ref{tab:main}, reinforcing the claim that quantum reinforcement learning enhances both fidelity and perceptual realism.

\bibliographystyle{ieeetr}
\bibliography{ref}

\end{document}